\documentclass[aps,prb,superscriptaddress,twocolumn,floatfix,showpacs]{revtex4}
\usepackage{graphicx}
\usepackage[]{amsmath}

\begin{document}
\newcommand{\chem}[1]{\ensuremath{\mathrm{#1}}}

\title{Thermodynamic and magnetic properties of the 
layered triangular magnet \chem{NaNiO_2}}
\author{P.\ J.\ Baker}
\affiliation{Clarendon Laboratory, University of Oxford, 
Parks Road, Oxford OX1
3PU, United Kingdom}

\author{T.\ Lancaster}
\affiliation{Clarendon Laboratory, University of Oxford, 
Parks Road, Oxford OX1
3PU, United Kingdom}

\author{S.\ J.\ Blundell}
\affiliation{Clarendon Laboratory, University of Oxford, 
Parks Road, Oxford OX1
3PU, United Kingdom}

\author{M.\ L.\ Brooks}
\affiliation{Clarendon Laboratory, University of Oxford, 
Parks Road, Oxford OX1
3PU, United Kingdom}

\author{W.\ Hayes}
\affiliation{Clarendon Laboratory, University of Oxford, 
Parks Road, Oxford OX1
3PU, United Kingdom}

\author{D.\ Prabhakaran}
\affiliation{Clarendon Laboratory, University of Oxford, 
Parks Road, Oxford OX1
3PU, United Kingdom}

\author{F.\ L.\ Pratt}
\affiliation{ISIS Muon Facility, ISIS, Chilton, Oxon. 
OX11 0QX, United Kingdom}

\date{\today}

\begin{abstract}
We report muon-spin rotation, heat capacity, magnetization, 
and ac magnetic susceptibility measurements of the magnetic 
properties of the layered spin-$1/2$ antiferromagnet \chem{NaNiO_2}.
These show the onset of long-range magnetic order below 
$T_{\rm N} = 19.5\,{\mathrm K}$.
Rapid muon depolarization, persisting from $T_{\rm N}$ to 
about $5\,{\mathrm K}$ above $T_{\rm N}$, is consistent with 
the presence of short-range magnetic order.
The temperature and frequency dependence of the ac susceptibility 
suggests that magnetic clusters persist above $25\,{\mathrm K}$ 
and that their volume fraction decreases with increasing temperature.
A frequency dependent peak in the ac magnetic susceptibility at 
$T_{\rm sf} = 3\,{\mathrm K}$ is observed, consistent with a 
slowing of spin fluctuations at this temperature.
A partial magnetic phase diagram is deduced.
\end{abstract}


\maketitle

Geometrically frustrated transition metal oxides exhibit a 
rich variety of magnetic behavior due to competing interactions. 
A significant degeneracy in the ground state leads to such 
phenomena as spin glass \cite{gardner83}, spin liquid 
\cite{mila00}, and spin ice \cite{snyder01} phases.
Triangular lattice antiferromagnets exhibit a variety of 
these phenomena \cite{collins97}.
When the triangles forming the lattice are distorted from 
equilateral to isosceles there is a partial release of the 
geometrical frustration, which can lead to more unusual forms 
of magnetic order \cite{coldea03, kobayashi99, sales04}.
Among triangular lattice antiferromagnets, \chem{LiNiO_2} 
\cite{chatterji05}, \chem{AgNiO_2} \cite{kikuchi99}, and 
\chem{NaNiO_2} have somewhat enigmatic magnetic behavior.
The difficulty of producing stoichiometric \chem{LiNiO_2} 
has led to a variety of sample-dependent results \cite{nunez00}.
\chem{AgNiO_2} can be produced in stoichiometric form but 
no magnetic Bragg peaks have so far been reported \cite{kikuchi99}.
Recently neutron powder diffraction studies have determined 
the low temperature magnetic structure of \chem{NaNiO_2} 
\cite{darie05,lewis2004}.
\textcite{darie05} find the ordering of the magnetic moments 
at $4\,{\mathrm K}$ to be a slight modification of the A-type 
antiferromagnetic ordering previously proposed \cite{bongers66}.
The magnetic moments were found to be aligned at an angle of 
$100(2)^{\circ}$ to the $a$-axis in the $ac$ plane with no 
moment along the $b$-axis.
A peak in the magnetic susceptibility interpreted as the 
N\'eel temperature, $T_{\rm N}$, has been observed around 
$20\,{\mathrm K}$ \cite{bongers66, kemp90, chappel609}.
The Curie-Weiss constant, $\theta_{\rm CW} = +36\,{\mathrm K}$
~\cite{chappel609}, shows the presence of ferromagnetic interactions 
above $T_{\rm N}$.

The intralayer and interlayer exchange constants of \chem{NaNiO_2}, 
$J_{\parallel} = -13.3\,{\mathrm K}$ and $J_{\perp} = 1.3\,{\mathrm K}$, 
were determined from a model assuming an A-type antiferromagnetic 
ordering with an anisotropy field \cite{holzapfel04}; the layers 
are sufficiently strongly coupled to permit long range magnetic 
order below $T_{\rm N}$.
The Ni-O-Ni bond angles are $\approx 95^\circ$ at room temperature 
\cite{chappel615}. 
An undistorted $90^\circ$ geometry favours weak ferromagnetic 
superexchange, while a large deviation from a 90$^\circ$ bond 
angle can reverse the sign of this exchange coupling \cite{tornow99}. 
In NaNiO$_2$ it appears that despite the distortion, in-plane 
ferromagnetic coupling prevails, though the precise nature of 
the spin and orbital ordering remains under discussion 
\cite{dare03, vernay04, reitsma05}.

Above $480\,{\mathrm K}$ the space group of \chem{NaNiO_2} 
is rhombohedral ($R \bar{3} m$) and there is a cooperative 
Jahn-Teller transition to a low temperature monoclinic ($C 2 / m $) 
phase below this temperature \cite{chappel615}.
The low-temperature structure can be considered to be layers of 
\chem{NiO_6} octahedra in the $ab$ plane, with a trigonal 
distortion lengthening the \chem{Ni}-\chem{O} bonds in the 
$ac$ plane along an axis at $41^{\circ}$ to the $c$-axis of 
the crystal.
The \chem{Ni^{3+}} ($3d^7$) ion is in the low spin state 
($t^{6}_{2g} e^{1}_{g}$, $S = 1/2$), so the ground state 
is a singly-occupied $\vert 3 z^2 - r^2 \rangle$ orbital 
with $z$ along the axis of the Jahn-Teller induced trigonal 
distortion \cite{chappel615}.

In this paper we present the results of zero-field muon spin 
rotation ($\mu$SR) experiments \cite{blundell99} on polycrystalline 
\chem{NaNiO_2} together with heat capacity and magnetic 
susceptibility measurements.
These provide evidence of long range magnetic order below 
$T_{\rm N}$, a slowing of spin fluctuations at low temperature, 
and short range magnetic order within a small temperature region 
above $T_{\rm N}$, with magnetic correlations persisting to 
higher temperature.

\begin{figure}[ht]
\includegraphics[width=8.5cm]{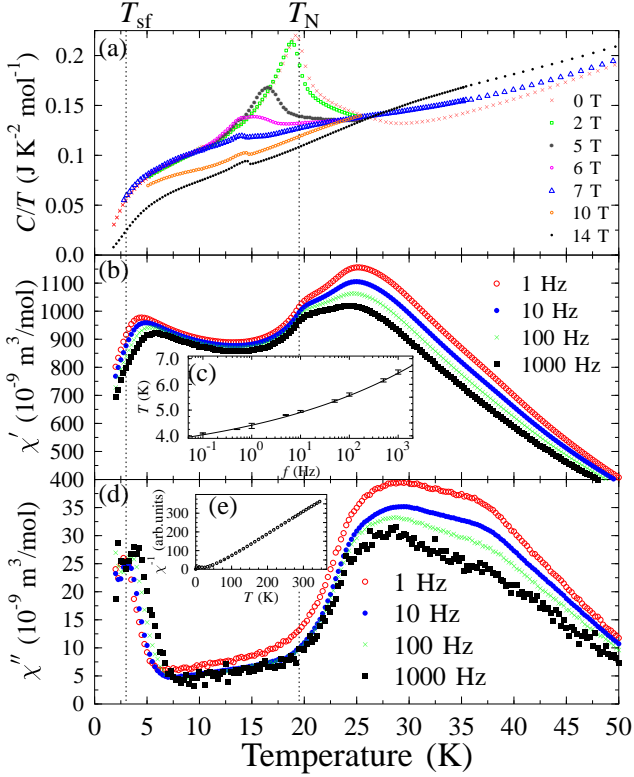}
\caption{(Color online) The panels correspond to: 
(a) Heat capacity divided by temperature in fields between $0$ 
and $14\,{\mathrm T}$. 
(b) Real part, $\chi^{\prime}$, of the ac magnetic susceptibility.
(c) Temperature dependence of the peak in $\chi^{\prime}$ associated 
with $T_f$ with a fit to the Ogielski relation (Eq.~(\ref{ogielski})).
(d) Imaginary part, $\chi^{\prime\prime}$, of the ac magnetic susceptibility.
(e) Inverse of magnetic susceptibility data against temperature 
with a linear fit showing the high-temperature Curie-Weiss behavior.
The vertical dashed lines indicate temperatures referred to in the text.}
\label{labstack}
\end{figure}

\chem{NaNiO_2} was prepared from \chem{Na_2 O_2} and \chem{NiO} 
powders heated at $700^{\circ} \mathrm C$ for 100 hours under 
pure oxygen flow, with intermediate grinding.
X-ray powder diffraction showed that the impurity concentration 
was below the $2\,\%$ resolution limit of the apparatus.
Heat capacity data measured in magnetic fields between $0$ and 
$14\,{\mathrm T}$, taken with a Quantum Design PPMS, are shown 
in Fig.~\ref{labstack}(a).
In zero field the transition at $T_{\rm N}=19.5\,{\mathrm K}$ 
is seen as a rather broad peak, and there is no evidence for 
other phase transitions below $30\,{\mathrm K}$.
With increasing magnetic field the temperature of this peak 
decreases to about $14\,{\mathrm K}$ (see Fig.~\ref{labstack}(a)).

Magnetic susceptibility data, taken using a Quantum Design MPMS 
SQUID magnetometer are shown in Fig.~\ref{labstack}(b), (d) and (e).
The high temperature dc susceptibility data, shown in 
Fig.~\ref{labstack}(e), are consistent with 
$\theta_{\rm {CW}} = +36\,{\mathrm K}$ \cite{chappel609}.
The real and imaginary parts of the ac susceptibility, 
$\chi^{\prime}$ and $\chi^{\prime\prime}$, are presented in 
Figs.~\ref{labstack}(b) and (d) (driving field $3.5\,{\mathrm {Oe}}$).
They show a frequency dependent peak in $\chi^{\prime}$ located 
slightly above a spin-freezing temperature $T_{\rm sf}$, which is 
determined below.
Figure~\ref{labstack}(d) shows the temperature variation of this 
peak in $\chi^{\prime}$ with frequency.
It was found that this could be fitted to the Ogielski scaling 
relation \cite{ogielski85}:
\begin{equation}
T = T_{\rm sf} (1 + (f \tau_0)^{1/{z \nu_c}}),
\label{ogielski}
\end{equation}
where $T$ is the temperature of the peak in $\chi^{\prime}$, $f$ 
is the measurement frequency, $\tau_0$ is the relaxation time of the 
system, $z$ is a dynamic exponent, and $\nu_c$ is a critical exponent.
Fitting the data to Eq.~(\ref{ogielski}) gives 
$T_{\rm sf} = 3 \pm 0.2\,{\mathrm K}$, 
$\tau_0 = 5.4(2) \times 10^{-3}\,{\mathrm {s}}$ and $z \nu_c = 8.1 \pm 0.4$.
This is typical of glassy behavior or the slowing of spin fluctuations.
Given that muon precession was observed down to $1.6\,{\mathrm K}$ 
(see below), the formation of a true spin glass around $T_{\rm sf}$ 
is excluded.
One possible interpretation of the feature around $T_{\rm sf}$ 
is that it could arise from a small concentration of defects 
comprised of pairs of \chem{Ni^{2+}} impurity spins which are 
associated with oxygen vacancies.
Alternatively, the fluctuations in the spins could be slowing 
down around $T_{\rm sf}$.
At $T_{\rm N}$, $\chi^{\prime}$ has a small maximum and rises 
to a larger peak near $25\,{\mathrm K}$, with the frequency 
dependence increasing from $T_{\rm N}$ to the peak, and decreasing above it.
The temperature of the peak in $\chi^{\prime}$ at $25\,{\mathrm K}$ 
decreases slowly with increasing frequency, which may be related 
to the presence of two sets of relaxation times varying differently 
with temperature (see below).
$\chi^{\prime\prime}$ also rises sharply near $25\,{\mathrm K}$ but 
to a plateau continuing up to $\sim 35\,{\mathrm K}$.
Together these suggest that short-range order persists up to 
$\sim 25\,{\mathrm K}$, that slowly fluctuating clusters of spins 
are present within a fast fluctuating paramagnetic bulk above this 
temperature, and that the volume fraction of clusters decreases with 
increasing temperature.

\begin{figure}[t]
\includegraphics[width=8.5cm]{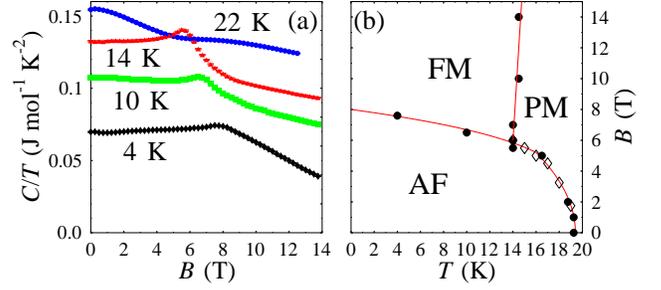}
\caption{(Color online) 
(a) Heat capacity divided by temperature vs field at four temperatures.
(b) Partial magnetic phase diagram deduced from heat capacity 
($\bullet$) and magnetization ($\diamond$) data.
AF: A-type antiferromagnetic phase.
PM: Paramagnetic phase.
FM: Ferromagnetic phase.
 \label{phases}}
\end{figure}

Constant temperature heat capacity data with varying magnetic field 
are presented in Fig.~\ref{phases}(a).
Except for the data taken at $22\,{\mathrm K}$, a peak is observed 
which corresponds to the field labelled $H_{{\mathrm C}1}$ in the 
magnetization data reported in Ref.~\onlinecite{holzapfel04}.
This suggests that this marks the upper field boundary of A-type 
antiferromagnetic order.
At $22\,{\mathrm K}$ the heat capacity decreases with increasing 
field consistent with short-range order.
The partial magnetic phase diagram deduced from our heat capacity 
and magnetization data is shown in Fig.~\ref{phases}(b).

Our zero-field $\mu$SR experiments were carried out using the 
DOLLY instrument at the Paul Scherrer Institute (PSI), Villigen, Switzerland.
In our $\mu$SR experiments, spin polarized positive muons 
($\mu^+$, mean lifetime $2.2\,\mu \mathrm s$, momentum 28~MeV$/c$) 
were implanted into polycrystalline \chem{NaNiO_2}. 
The decay positron asymmetry function, $A(t)$ \cite{blundell99}, 
is proportional to the average spin polarization of the muons stopped 
within the sample.
The muon spin precesses around an internal magnetic field, $B_{\mu}$, 
at a frequency $\nu_\mu = (\gamma_{\mu} / 2 \pi)\vert B_{\mu} \vert$, 
where $\gamma_{\mu}/2 \pi = 135.5$ MHz~T$^{-1}$.

The asymmetry data were fitted to Eq.~(\ref{fitfunc1})~\cite{ddr97} 
below $T_{\rm N}$, and to Eq.~(\ref{fitfunc2}) above $T_{\rm N}$:
\begin{eqnarray}
A(t) &=& A(0) \left( P_{1} e^{-\lambda_{1} t}
+ P_{2} e^{-\lambda_{2} t} \cos \left( 2 \pi \nu_{\mu} t + \phi_0 
\right) \right),
\label{fitfunc1} \\
A(t) &=& A(0) \left( P_{\rm f} e^{-\lambda_{\rm f} t} + P_{\rm s} 
e^{-\lambda_{\rm s} t} \right),
\label{fitfunc2}
\end{eqnarray}
where $A(0)$ is the initial asymmetry.
\begin{figure}[t]
\includegraphics[width=8.5cm]{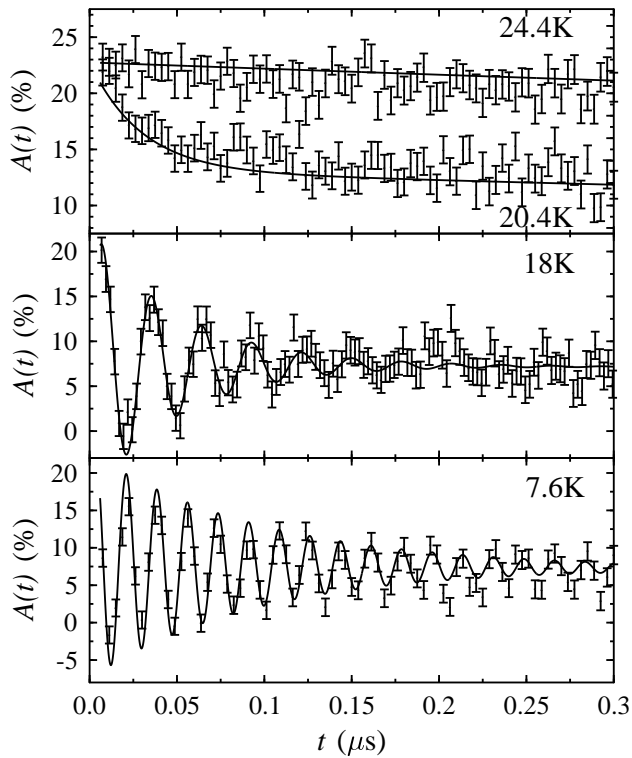}
\caption{Muon decay asymmetry in \chem{NaNiO_2} plotted at different 
temperatures. The solid lines are fits of the data to Eqs.~(\ref{fitfunc1})
~\&~(\ref{fitfunc2}) with the parameters shown in Fig.~\ref{stack}. 
\label{waterfall}}
\end{figure}
$P_{1}$ and $P_{2}$ are respectively the longitudinal and transverse 
components of the muon polarization, and $P_{1} + P_{2} = 1$.
The exponential relaxation associated with $P_{1}$ reflects the dynamical 
fluctuations of the fields being probed.
The $P_{2}$ term describes muon precession with a distribution of local 
fields dephasing the muon spins.
In a fully magnetically ordered polycrystalline sample we expect 
$P_{2}/P_{1}=2$.
Coherent muon precession will be observed if long range order is 
present within the sample.
$P_{\rm s}$ and $P_{\rm f}$ describe slow and fast dynamic fluctuations 
respectively.
A small initial phase offset, $\phi_0$, was observed below $T_{\rm N}$, 
larger than could be attributed to errors in determining the time that 
the muons enter the sample.
This could be produced by a small magnetic inequivalency in the position 
of muons stopped within the sample, consistent with an asymmetric peak 
seen in Fourier transforms of the data.
In the fitting procedure, data were fitted in the time range 
$0 < t < 8\,\mu s$, where the effect of background counts could be 
reliably subtracted.
Rapid dynamic fluctuations lead to $\lambda_{1} \propto {{\gamma_{\mu}}^2 
(\Delta B)^2}/ \nu$, where $\Delta B$ is the amplitude of the 
fluctuating local field and $\nu$ is the fluctuation rate \cite{ddr97}.

\begin{figure}[ht]
\includegraphics[width=8.5cm]{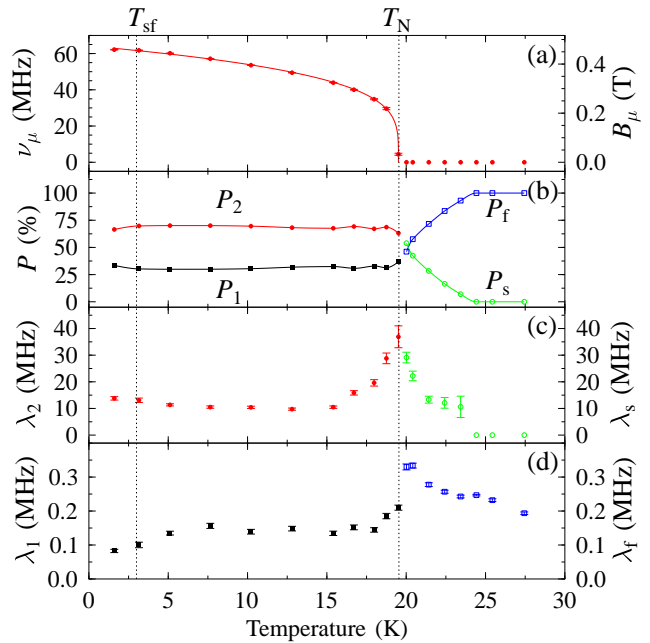}
\caption{
(Color online) Temperature dependence of the parameters determined 
from fitting data to Eqs.~(\ref{fitfunc1})~\&~(\ref{fitfunc2}): 
(a) the oscillation frequency, $\nu_{\mu}$, and the internal magnetic 
field, $B_{\mu}$, with a fit to Eq.~(\ref{nuoft}).
(b) Amplitudes of the relaxation components $P_1$ and $P_2$, and 
$P_{\rm f}$ and $P_{\rm s}$.
(c) Relaxation rates $\lambda_2$ and $\lambda_{\rm s}$.
(d) Relaxation rates $\lambda_1$ and $\lambda_{\rm f}$.
The vertical dashed lines indicate temperatures referred to in the text.
}
\label{stack}
\end{figure}

Spectra measured at four temperatures are shown in Fig.~\ref{waterfall}.
There are three distinct temperature regions apparent from the muon 
asymmetry spectra.
At low temperatures ($T \leq 19.5\,{\mathrm K}$) there are clear 
oscillations in the asymmetry showing that long range magnetic 
order exists and the observed ratio of $P_{2}:P_{1} \approx 2$ 
(see Fig.~\ref{stack}(b)) indicates that the sample is magnetic 
over its entire volume.
The value of $\lambda_{2,{\rm s}}$ is much larger than 
$\lambda_{1,{\rm f}}$ (see Fig.~\ref{stack}(c) \& (d)) so at short 
times only the effect of $\lambda_{2,{\rm s}}$ is seen in 
Fig.~\ref{waterfall}.
An intermediate temperature range ($19.5 < T < 24\,{\mathrm K}$) 
gives no oscillations, and the relaxation is modelled with the 
two exponential components of Eq.~(\ref{fitfunc2}), with the 
amplitude of the faster relaxing component decreasing with 
increasing temperature.
Above $24\,{\mathrm K}$ the relaxation is well described by a 
single exponential, $P_{\rm f} {\mathrm {exp}}(- \lambda_{\rm f} t)$, 
consistent with fast fluctuations of paramagnetic moments characterized 
by a single correlation time in the muon time window.

The temperature dependence of the parameters derived from fitting 
Eqs.~(\ref{fitfunc1})~\&~(\ref{fitfunc2}) to muon asymmetry spectra 
are presented in Fig.~\ref{stack}.
The muon precession frequency, $\nu_{\mu}$, in the ordered phase is 
shown in Fig.~\ref{stack}(a).
This is proportional to the sublattice magnetization at the muon site, 
and was fitted to a function \cite{borsa95}: 
\begin{equation}
\nu_{\mu}(T)=\nu_{\mu}(0)(1-(T/T_{\mathrm N}))^{\beta_{\rm m}}.
\label{nuoft}
\end{equation}
The fit gives $\nu_{\mu}(0) = 64.2(2)\,{\mathrm {MHz}}$ corresponding 
to a field at the muon site of $\sim 0.5\,{\mathrm T}$.
Dipole field calculations show that this field will be experienced by 
muons near any of the oxygen atoms in the octahedron surrounding a 
nickel atom, in regions of high electron density \cite{meskine04}, 
and show that our results are consistent with the magnetic structure 
determined by \textcite{darie05}.
Our calculations also suggest that the muon precession frequency is 
insensitive to small deviations from this magnetic structure.
Fitting Eq.~(\ref{nuoft}) to the muon precession frequencies gave 
$T_{\rm N} = 19.51(1)\,{\mathrm K}$ and $\beta_{\rm m} = 0.24(1)$.
This value of $\beta_{\rm m}$ suggests that the system is behaving 
as a 2D XY magnet \cite{bramwell95}.

In relation to the peak just above $T_{\rm sf}$ in the magnetic 
susceptibility, we note that extrapolating the Ogielski scaling 
relation to the muon time window suggests a maximum in the dynamic 
relaxation rate $\lambda_1$ should be observed around $7\,{\mathrm K}$, 
and a broad maximum of low amplitude is just detectable at this 
temperature (Fig.~\ref{stack}(d)).
The presence of two exponential relaxation components above 
$T_{\rm N}$ (see Eq.~(\ref{fitfunc2})) suggests that short-range 
magnetic order persists over a small temperature range of 
$\sim 5\,{\mathrm K}$ above $T_{\rm N}$.
The fluctuations in the magnetic field producing the faster relaxing 
component, $\lambda_{\rm s}$, are two orders of magnitude slower than 
those producing the slowly relaxing component, $\lambda_{\rm f}$.
$P_{\rm s}$ decreases with increasing temperature up to $24\,{\mathrm K}$, 
showing that the ratio of slow to fast dynamic relaxation is decreasing.
Above this temperature the muon relaxation is that expected for a system 
in the fast-fluctuation regime.
The slow spin relaxations observed in the frequency dependence of the 
ac susceptibility above $25\,{\mathrm K}$ are not within the muon time 
window so are not observed.
The observation of two components in the muon relaxation below 
$24\,{\mathrm K}$, together with a changing frequency dependence 
of the ac susceptibility, suggest a model of coalescing magnetic 
clusters forming well above $T_{\rm N}$.
On cooling below $\sim 50\,{\mathrm K}$, these magnetic clusters 
occupy an increasing volume fraction, causing the increase in the 
frequency dependence of the ac susceptibility, until at 
$\sim 25\,{\mathrm K}$ the sample possesses short-range 
antiferromagnetic order.
Below $25\,{\mathrm K}$ these clusters coalesce, leading to the 
decrease in the frequency dependence of the ac susceptibility, 
and the increase in $P_{\rm s}$, until at $T_{\rm N}$ the sample 
possesses long-range order.
Short-range order within the layers and antiferromagnetically correlated 
layers is consistent with the persistence of unbroadened magnetic Bragg 
peaks observed by neutron powder diffraction over a similar temperature 
range above $T_{\rm N}$\cite{lewis2004, gaulin05}.

In conclusion, \chem{NaNiO_2} shows the onset of long-range magnetic 
order at $T_{\rm N} = 19.5\,{\mathrm K}$, with the dependence of the 
sublattice magnetization on temperature appropriate for a 2D XY magnet.
The slowing of spin fluctuations above $T_{\rm sf}$ is evident in the 
ac magnetic susceptibily data.
At temperatures just above $T_{\rm N}$ there is evidence of short-range 
order, and of magnetic clusters persisting within a paramagnetic phase 
above this temperature.

Part of this work was performed at the Swiss Muon Source, Paul Scherrer
Institute, Villigen, Switzerland. 
We are grateful to Robert Scheuermann for experimental assistance, and 
to M.\,Holzapfel, B.\,D.\,Gaulin, and A.\,Coldea for helpful discussions.
T.\ L.\ acknowledges support from the European Commission under the 
$6^{\rm {th}}$ Framework Programme through the Key Action: Strengthening 
the European Research Area, Research Infrastructures.
Contract no: RII3-CT2003-505925.
This work was funded by the EPSRC (UK).

\bibliography{NaNiO2v5}

\end{document}